\documentclass[11pt,a4paper]{article}
\usepackage[hyperref]{emnlp2020}
\usepackage{times}

\usepackage{xcolor}
\usepackage{url}
\usepackage{amsmath, multirow}
\usepackage{latexsym, soul}
\usepackage[inline]{enumitem}
\usepackage{graphicx, float}
\usepackage{array}
\usepackage{tabu}
\usepackage{booktabs}
\usepackage{xspace,mfirstuc,tabulary}

\usepackage{microtype}

\aclfinalcopy 

\title{End-to-End Speech Recognition and Disfluency Removal}

\author{Paria Jamshid Lou$^1$ and Mark Johnson$^{2, 1}$\\
  $^1$Department of Computing, Macquarie University \\
  $^2$Oracle Digital Assistant, Oracle Corporation\\
  \texttt{$^1$paria.jamshid-lou@hdr.mq.edu.au}\\
  \texttt{$^2$mark.mj.johnson@oracle.com} \\}
\date{}

\definecolor{editcolor}{rgb}{0.10, 0.56, 0.6}
\definecolor{intcolor}{rgb}{0.65, 0.74, 0.85}

\begin{document}
\maketitle
\begin{abstract} 
Disfluency detection is usually an intermediate step between an automatic speech recognition (ASR) system and a downstream task. By contrast, this paper aims to investigate the task of end-to-end speech recognition and disfluency removal. We specifically explore whether it is possible to train an ASR model to directly map disfluent speech into fluent transcripts, without relying on a separate disfluency detection model. We show that end-to-end models do learn to directly generate fluent transcripts; however, their performance is slightly worse than a baseline pipeline approach consisting of an ASR system and a specialized disfluency detection model. We also propose two new metrics for evaluating integrated ASR and disfluency removal models. The findings of this paper can serve as a benchmark for further research on the task of end-to-end speech recognition and disfluency removal in the future.
\end{abstract}

%
%
\section{Introduction}
Disfluency is a characteristic of spontaneous speech which is not present in written texts. Disfluencies include filled pauses (e.g.~\emph{um} and \emph{uh}), repetitions (e.g.~\emph{the the}), corrections (e.g.~\emph{Show me the flights~$\dots$~the early flights}), parenthetical asides (e.g.~\emph{you know}), interjections (e.g.~\emph{well} and \emph{like}), restarts (e.g.~\emph{There's a}~$\dots$~\emph{Let's go}) and  partial words (e.g.~\emph{wou-} and \emph{oper-}) which frequently occur in spontaneous speech\footnote{\citet{shri:94} observed disfluencies once in every 20 words.} and reduce the readability of speech transcripts~\citep{liu:06}. They also pose a major challenge to downstream tasks relying on the output of speech recognition systems, such as parsing and machine translation models~\citep{john:04, wang:10, hon:14}. Since these models are usually trained on fluent clean corpora, the mismatch between the training data and the actual use case decreases their performance. To tackle this challenge of spontaneous speech, specialized disfluency detection models are developed and applied as a post-processing step to remove disfluencies from the output of speech recognition systems~\citep{zay:16, wang:18, dong:19}. 

One type of disfluency which is especially problematic for disfluency detection models is speech repair. \citet{shri:94} defines three distinct parts of a speech repair, referred to as \emph{reparandum}, \emph{interregnum} and \emph{repair}. As illustrated in the example below, the reparandum \emph{to Boston} is the part of the utterance that is replaced and is usually followed by an interruption point in the speech signal, the interregnum \emph{uh I mean} is an optional part of a disfluent structure (that consists of a filled pause \emph{uh} and a discourse marker \emph{I mean}) and the repair \emph{to Denver} replaces the reparandum. The fluent version is obtained by deleting reparandum and interregnum words. 


\begin{equation*}  \label{ex:0}
\centering 
\begin{array}{l}
\mbox{\textit{I want a flight}} \underbrace{\mbox{\it\strut \textcolor{editcolor}{to Boston}}}_{\mbox{\scriptsize  \textcolor{editcolor}{reparandum}}}\strut\hspace{-0.5cm}~~~~~\underbrace{\mbox{\it \textcolor{intcolor}{uh I mean}\strut}}_{\mbox{\scriptsize  \textcolor{intcolor}{interregnum}}} 
\underbrace{\mbox{\it\strut to Denver.}}_{\mbox{\scriptsize repair}}
\end{array}
\end{equation*}

Disfluency detection is usually an intermediate step between an ASR model and a downstream task. This pipeline approach is complex to implement and leads to higher inference latency. It also has the potential problem of errors compounding between components, e.g. recognition errors lead to larger disfluency detection errors. End-to-end models, on the other hand, are less prone to such problems. More importantly, end-to-end models can leverage paralinguistic features in speech signal that are not available in pipeline systems. Speech carries extra information beyond the words which might provide useful cues to disfluency detection\footnote{Prosodic cues (e.g. pause) signal disfluencies by marking the interruption point~\citep{shri:94, zay:19}.}. In this paper, we address the task of end-to-end speech recognition and disfluency removal. Specifically, we investigate whether it is possible to train an ASR model end-to-end to directly map disfluent speech into fluent transcripts, without an intermediate disfluency detection step. Some previous work has attempted disfluency detection as part of another task in an end-to-end manner, e.g. joint disfluency detection and constituency parsing~\citep{jam:19} and direct translation from disfluent Spanish speech to fluent English transcripts~\citep{sal:19}. However, to the best of our knowledge, this is the first work that systematically investigates the task of end-to-end ASR and disfluency removal, serving as a starting point for future research into end-to-end disfluency removal systems. In this paper, we aim to answer the following questions:
\begin{itemize}
\item \emph{Can an ASR model directly generate fluent transcripts from disfluent speech?} We might expect an end-to-end ASR model (without an explicit disfluency detection component) not to effectively detect disfluencies. However, we show that \textbf{end-to-end ASR models do learn to directly generate fluent transcripts} and their performance is comparable to a baseline pipeline system (i.e. an ASR model followed by a specialized disfluency detection model).




\item \emph{How does the choice of architecture impact disfluency detection and removal in end-to-end speech recognition?} We compare the performance of three neural-based end-to-end ASR and disfluency removal models including a Connectionist Temporal Classification based model, an LSTM-based sequence-to-sequence model and a Transformer sequence-to-sequence model and show that \textbf{a Transformer ASR model has the best performance} on disfluency removal.

\item \emph{How can we systematically evaluate the performance of an end-to-end ASR and disfluency removal model?} The existing evaluation metrics are designed to measure the performance of a single task, namely speech recognition or disfluency detection, but not both. We introduce \textbf{two new metrics measuring the disfluency removal and word recognition performance} of an end-to-end model. 
\end{itemize}
 
\section{Related Work}
Disfluency removal is typically performed by training a specialized disfluency detection model on disfluency labeled data and applying it as a separate component following an ASR model and prior to a downstream task. The specialized disfluency detectors~\citep{zay:16, wang:16, jam:18} are usually trained on the Switchboard corpus~\citep{mit:99} which is the largest available dataset with gold (i.e. human-annotated) disfluency labels. State-of-the-art disfluency detectors use Transformer models with pretrained contextualised word embeddings (e.g.~BERT)~\citep{tran:19, jam:19, dong:19, wang:19, jam:20}. Multi-task learning has been effective for disfluency detection, for example, a Transformer trained to jointly detect disfluencies and find constituency parse trees would leverage syntactic information and detect disfluencies more accurately~\citep{jam:19}. Self-training and ensembling have also shown to provide benefit to disfluency detection~\citep{jam:20}. Self-training on disfluent data provides benefits orthogonal to the pretrained contextualized embeddings and mitigates the scarcity of gold disfluency labeled data. The BERT-based self-attentive parser introduced in~\citet{jam:19} is the current state-of-the-art in disfluency detection; thus, we use it as the ``off-the-shelf'' disfluency detector in our pipeline approach, as explained in Section~\ref{sec:ex}.


With the rise of end-to-end models, the conversational speech translation models that directly translate disfluent speech into fluent texts have recently attracted increasing attention~\citep{sal:19, ansari:20, fukuda:20, saini:20}. The most similar previous work to ours is~\citet{sal:19}. They train a sequence-to-sequence model (called fluent model) to directly translate from disfluent Spanish speech to fluent English transcripts without a separate disfluency detection step. As a baseline, they train a model (called disfluent model) on disfluent speech and disfluent translations. To compare the performance of the fluent and disfluent models, they score the outputs against the fluent references using BLEU and METEOR. Similar METEOR scores are reported for both models, but BLEU scores are lower with the disfluent model. They argue that the disfluencies generated by the disfluent model lead to n-gram break-up in the fluent references and consequently decrease the BLEU scores. They conclude that higher BLEU scores in the fluent model imply that it is better at generating fluent translations. Although BLEU and METEOR are standard metrics for evaluating machine translation systems, they are not designed to evaluate the performance of end-to-end models in terms of disfluency removal. Since many disfluent words are copies of fluent words, the BLEU score of disfluent transcripts (e.g.~\emph{The the snack was delicious}) can be higher than that of fluent transcripts containing translation errors (e.g.~\emph{The meal was delicious} against the fluent reference \emph{The snack was delicious}). Furthermore, these metrics are sensitive to sequence length which makes them undesirable for evaluating end-to-end models incorporating disfluency removal. Fluent transcripts tend to contain fewer tokens per sentence in comparison with disfluent transcripts. By contrast, we introduce two new metrics in this paper that systematically measure the fluency of the generated transcripts. We also benchmark our end-to-end model against a state-of-the-art pipeline approach to explicitly evaluate its disfluency detection performance.

\section{Speech Recognition and Disfluency Removal Models}
We investigate three different ASR architectures: Connectionist Temporal Classification (CTC), LSTM-based and Transformer sequence-to-sequence models. Each of these three ASR models is trained twice: \begin{enumerate*}[label=(\roman*)] \item in a pipeline approach where the ASR model is trained to transcribe speech, followed by an ``off-the-shelf'' specialized disfluency detection model, \item in an end-to-end approach where the ASR model is trained to jointly transcribe speech and remove disfluencies, which we refer to as an \emph{integrated ASR and disfluency model}.\end{enumerate*} The ASR models for the two training regimes are identical in terms of architecture and the number of parameters. The only difference is their training data, i.e. the pipeline ASR model is trained on disfluent speech and disfluent transcripts while the end-to-end ASR model is trained on disfluent speech and fluent transcripts. Given the same speech utterance, the same ASR architecture is trained to either produce (i) or (ii):

\begin{enumerate}[label=(\roman*)]
\item \emph{I want a flight to Boston uh I mean to Denver}
\item \emph{I want a flight to Denver}
\end{enumerate}
 
As input features to the ASR model, we preprocess the speech signal by sampling the raw audio waveform using a sliding window of 25ms with stride 10ms. We extract 80-dimensional log mel-filterbank coefficients plus three fundamental frequency features from the frames using Kaldi~\cite{kaldi:11}. We train a CTC-based ASR model, called Jasper~\cite{li:19}, using the OpenSeq2Seq Toolkit\footnote{\url{https://github.com/NVIDIA/OpenSeq2Seq}}~\cite{open:18}. Jasper contains 10 blocks of 1D-convolutional layers, each with 5 sub-blocks. A sub-block consists of a 1D-convolutional operation, batch normalization, clipped ReLU activation and dropout. There is a residual connection between each block which is added to the output of the last 1D-convolutional layer in the block before the clipped ReLU activation and dropout. The optimizer used to train the model is stochastic gradient descent with momentum and the loss is CTC~\cite{grav:06}. At decoding time, a candidate list is generated using word-level 4-gram language models and beam search with a width of 2048. For more details, see~\citet{li:19}. 

We build the encoder-decoder Sequence-to-Sequence model with Bahdanau attention~\cite{bahd:14} using the Espresso Toolkit\footnote{\url{https://github.com/freewym/espresso}}~\cite{wangg:19}. The Sequence-to-Sequence model uses a 4-layer 2D-convolution, followed by a 3-layer bidirectional LSTM as an encoder and a 3-layer LSTM as a decoder. We train the model using cross-entropy loss and an Adam optimizer. We leverage shallow fusion~\cite{gul:15} as a language model integration technique. The decoder with shallow fusion computes a weighted sum of two posterior distributions over subword units from the speech recognition model and from the neural language model. For more details, see~\citet{wangg:19}.
 
We also train a Transformer ASR model inspired by~\citet{moh:19} using the Fairseq Toolkit\footnote{\url{https://github.com/pytorch/fairseq}}~\cite{ott:19}. The Transformer replaces the sinusoidal positional embeddings at the encoder and the decoder with convolutional layers to capture the positional information. The encoder contains two 2D-convolutional blocks with layer norms and ReLU after each convolutional layer. Each convolutional block contains two convolutional layers followed by a 2D max pooling layer with kernel sizes of 3 and 2, respectively. The convolutional layers are used on top of 16 encoder transformer blocks with model hidden dimension 1024 and 16 attention heads. The decoder includes three 1D-convolutional layers, each with a kernel size of 3, and 6 decoder transformer blocks. The Transformer layers learn the global sequential structure of the input while the convolutional layers learn local relationships within a small context. The training criterion is cross-entropy loss and the model is optimized using adadelta. We employ shallow fusion and standard beam search with a beam size of 20 at decoding time. In order to have a fair comparison with other models, we do not pretrain the Transformer. For more details, see~\citet{ott:19}. The language models used in the three ASR models are extracted from or trained on the same data used for training the ASR models (i.e. on fluent transcripts for the end-to-end models and on disfluent transcripts for the pipeline models).







\section{Evaluating Integrated ASR and Disfluency Models}
%

The performance of ASR models is usually evaluated in terms of word error rate (WER). WER is calculated by finding an alignment between the reference transcript (which is human-transcribed speech) and ASR output so that a minimum number of edits (i.e. substitutions, insertions and deletions) are required for transcribing the ASR output to the reference transcript. Given an alignment, WER is the ratio between the number of incorrectly aligned words and the total number of words in the reference transcript:
\begin{equation}
WER = \dfrac{s+i+d}{n}
\end{equation}
where $s$, $i$ and $d$ are the number of substitutions, insertions and deletions and $n$ is the total number of words in the reference transcript. WER measures the overall word recognition performance without distinguishing between fluent and disfluent words. Since the reference transcript contains both fluent and disfluent words, a WER of zero on the full transcript means that the system returned all of the disfluent words as well as the fluent words, which is not what an integrated system should do\footnote{An integrated system is expected to recognize fluent words and discard disfluent words in the output.}. While WER with respect to the full reference transcript (containing both fluent and disfluent words) is not meaningful for integrated systems intended to produce fluent output, WER with respect to the fluent subsequence is a meaningful measure of overall system, since this is the intended output of an integrated system.  However, since disfluencies only comprise around 6\% of the total words, the WER score largely reflects how well fluent words are recognized, rather than how well the system handles disfluencies. A system may score poorly on WER even though it is perfect in terms of detecting disfluencies because it fails to correctly recognize the fluent words. 

Specialized disfluency detection models are usually evaluated using edited f-score. Edited f-score focuses more on detecting disfluent words, so it is a decent metric for highly skewed data like Switchboard. Calculating f-score, however, is not straightforward in end-to-end models as the model is expected to generate fluent outputs directly (rather than tagging disfluencies in the output).

To address the limitations of the existing metric, we introduce two new evaluation metrics\footnote{\url{https://github.com/pariajm/e2e-asr-and-disfluency-removal-evaluator}} which assess the output of an integrated model in terms of fluency and word recognition accuracy in Section~\ref{sub:fer}. We then demonstrate the problems associated with the standard ASR alignment algorithm and how it can lead to undesirable alignments for evaluating integrated ASR and disfluency models. As a solution, we modify standard alignment weights to correctly align reference transcripts (which may contain disfluencies) with integrated model outputs in Section~\ref{sub:align}.  

\begin{figure}[H]
\centering 
\includegraphics[width=0.48\textwidth]{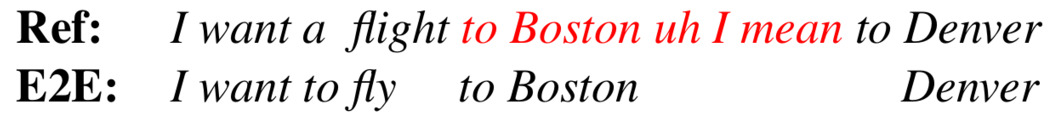}\caption{Ref is the reference transcript which is human-transcribed speech with gold disfluency labels, shown in red. E2E represents the output of an integrated ASR and disfluency removal model.}
\label{fig:00}
\end{figure}

\begin{figure*} [t]
\centering 
\includegraphics[width=0.9\textwidth]{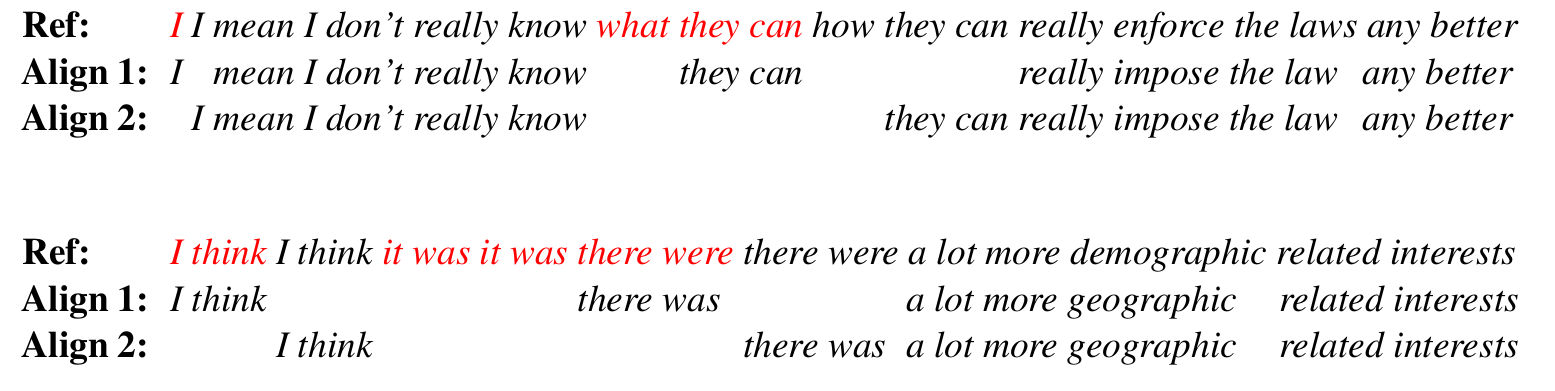}\caption{Ref is the reference transcript which is human-transcribed speech with gold disfluency labels, shown in red. Align 1 represents the alignment between the output of an integrated ASR and disfluency model and the reference transcript generated by the standard alignment weights where equal costs are allocated for aligning fluent and disfluent words. Align 2 is the desired alignment in order to make meaningful FER and DER evaluations. }
\label{fig:01}
\end{figure*}


\subsection{Fluent and Disfluent Error Rate Scores} \label{sub:fer}



To overcome the limitations of WER, we use the standard WER evaluation to evaluate fluent and disfluent words separately. In this way, the quality of integrated model outputs is evaluated in terms of both fluency and word recognition. We calculate the word error rate on fluent words (which we call the fluent error rate or FER) as the number of substitutions $s_f$, deletions $d_f$ and insertions $i_f$ among fluent words divided by the total number of fluent words $n_f$ in the reference transcript as below. For example, FER is equal to 0.5 in Figure~\ref{fig:00}, which is calculated on the fluent subset words shown in black where $s_{f}=2, i_{f}=0, d_{f}=1$ and $n_{f}=6$.
\begin{equation}
FER = \dfrac{s_{f}+i_{f}+d_{f}}{n_{f}}
\end{equation}

We define the word error rate on disfluent words (which we call the disfluent error rate or DER) as anything other than a deletion (i.e. substitutions $s_d$, insertions $i_d$ and copies $c_d$) among disfluent words divided by the total number of disfluent words $n_d$ in the reference transcript as below. For instance, DER is equal to 0.4 in Figure~\ref{fig:00}, which is calculated on the disfluent subset words shown in red where  $s_{d}=0, i_{d}=0, c_{d}=2,$ and $n_{d}=5$.
\begin{equation}
DER = \dfrac{s_{d}+i_{d}+c_{d}}{n_{d}}
\end{equation}

For calculating FER and DER, we need to align the reference transcripts (i.e. human-transcribed speech with gold disfluency labels) to the integrated model outputs, which are expected to be fluent. The aligner used for this purpose is explained in the following section.

\subsection{Aligning Integrated Model Output to Reference Transcripts} \label{sub:align}
In this section, we first describe the standard ASR alignment algorithm and explain why it sometimes finds misleading alignments of the output from integrated ASR and disfluency systems. We then suggest a modification to the standard edit distance alignment weights so that they lead to meaningful alignments between the reference transcript and the integrated model output.

\subsubsection{Problems with Standard ASR Alignment} \label{sec:prob}

To illustrate the problems with standard ASR alignment algorithms, consider Figure~\ref{fig:01}, where the outputs from an integrated model have been aligned with the reference transcripts using two different alignment weights. The first alignment, indicated as Align 1, is generated by the Sclite Toolkit. Sclite\footnote{\url{https://github.com/usnistgov/SCTK}} is a standard toolkit for evaluating ASR outputs which finds an alignment using dynamic programming algorithms such that a copy, deletion, insertion and substitution cost 0, 3, 3 and 4, respectively. Align 2, on the other hand, is what we expect an aligner to produce in order to have meaningful FER and DER evaluations. As shown in Align 1, the fluent words in the outputs of the integrated system are aligned with the disfluent words in the reference transcripts rather than the fluent words. Since we expect the reference transcript to contain both fluent and disfluent words and the output of an integrated system to discard the disfluencies, the standard alignment weights fail to properly align the integrated model output to the reference transcript. Align 1 and Align 2 have the same alignment cost with the standard weights, so an aligner using the standard weights has no reason to prefer one over the other. The problem that arises here is that since many disfluent words are copies of fluent words, if the same cost is used to align fluent and disfluent words, the alignment will be ambiguous (i.e. there will be multiple alignments with the same cost). Thus, to force the aligner to prefer aligning null (i.e. deletions) for disfluent words and copy for fluent words, we modify the alignment weights so the intuitively correct alignment scores better, and so will be chosen by the alignment algorithm.

\begin{figure*} [t]
\centering 
\includegraphics[width=0.9\textwidth]{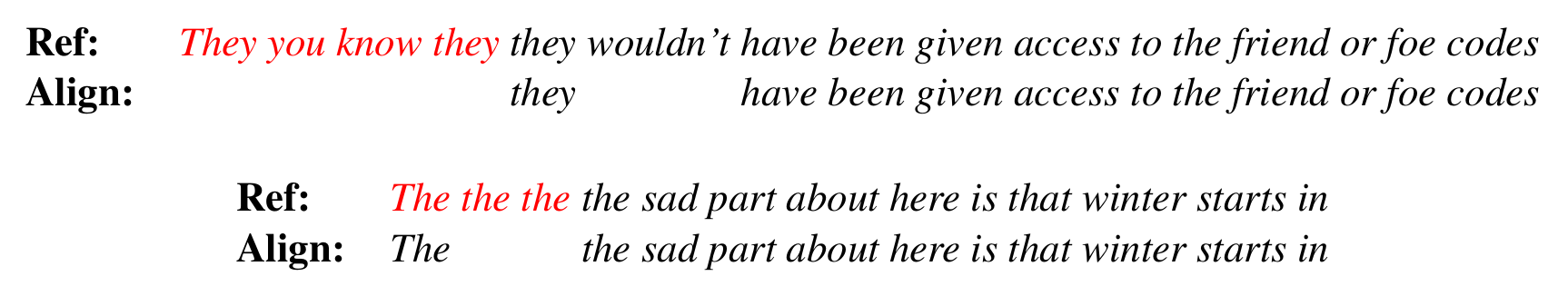}\caption{Ref is the reference transcript which is human-transcribed speech with gold disfluency labels, shown in red. Align refers to the alignment between the integrated ASR and disfluency model output and the reference transcript generated by the modified alignment weights where different costs are allocated for aligning fluent and disfluent words. }
\label{fig:02}
\end{figure*}  

\subsubsection{Alignment Weights for Integrated ASR and Disfluency Models}
We use two sets of weights for finding an alignment between the reference and the integrated model output. We use the standard alignment weights described in Section~\ref{sec:prob} for aligning fluent words, and slightly modify the weights to discourage aligning disfluent words in the reference transcript with words in the integrated model output. For the fluent region, a correct alignment operation is a \emph{copy} while for the disfluent region, a correct alignment is a \emph{deletion}. As shown in Table~\ref{tab:01}, the alignment cost is slightly higher for inserting, copying and substituting a disfluent word and slightly lower for deleting a disfluent word. Having a higher alignment cost for disfluent words results in a preference to align the words in integrated model outputs with fluent words as illustrated in Figure~\ref{fig:02}. Ambiguities can still arise even if disfluent words have a higher alignment cost than fluent words. However, these ambiguities do not affect the disfluency evaluation scores as our disfluency evaluation scores only depend on whether a word is disfluent or not.

\begin{table}[H]
\begin{center}
\begin{tabular}{lcc}
	\midrule \bf Operation & \bf Fluent & \bf Disfluent\\ 	\midrule
Copy (c) & $0$ & $0+10^{-7}$ \\		
Insertion (i) & $3$ & $3+10^{-7}$\\ 
Deletion (d)  & $3$ & $3-10^{-7}$\\ 			
Substitution (s)   & $4$ & $4+10^{-7}$\\ 	\midrule		
\end{tabular}
\end{center}
\caption{The two sets of weights used to align disfluent and fluent words separately. } \label{tab:01} 
\end{table}
In summary, although WER is a standard metric for evaluating ASR models, it is insufficient for evaluating integrated ASR and disfluency systems as it measures the overall word recognition accuracy, and does not specifically focus on how well the end-to-end system handles disfluencies. Alternatively, we propose a modified alignment strategy with different weights for fluent and disfluent word alignments. Thus, it is possible to calculate word error rate on fluent and disfluent regions separately. Our new evaluation metrics and alignment weights are useful for aligning and evaluating any system trained to remove disfluency in its output.

\begin{table*}[t]
\begin{center}
\begin{tabular}{p{0.2cm}p{0.99cm}p{13.1cm}}
  \midrule

\multirow{3}{*}{\bf 1} & \bf{Ref} & \emph{$\dots$ the rights of that individual \textcolor{red}{are been} have been you know impugned $\dots$}\\
& \bf{Pipe} & \emph{$\dots$ the rights of that individual or\hphantom{e} ben\hphantom{e} have been you know immune\hphantom{d} $\dots$}\\ 

& \bf{E2E} & \emph{$\dots$ the rights of that individual \hphantom{are been} have been you know impuned\hphantom{e} $\dots$}\\

\midrule

\multirow{3}{*}{\bf 2} & \bf{Ref} & \emph{$\dots$ \textcolor{red}{I} actually \textcolor{red}{my dad's} \hphantom{i}~my dad's almost ninety $\dots$}\\

& \bf{Pipe} & \emph{$\dots$ I yeah\hphantom{ally} \hphantom{my} cause  my dad's almost ninety $\dots$}\\

& {\bf E2E} & \emph{$\dots$ \hphantom{I} actually \hphantom{my dad'se}~my dad's almost ninety $\dots$}\\
 \midrule


\multirow{3}{*}{\bf 3} & \bf{Ref} & \emph{\textcolor{red}{I've been to\hphantom{nt} a couple o-} I've been to a few games before}\\

& \bf{Pipe} & \emph{\hphantom{I've} I\hphantom{een} bent a couple \hphantom{o-} I've been to a few games before}\\

& {\bf E2E} & \emph{\hphantom{I've been to\hphantom{nt} a couple o-} I've been to a few games before}\\
\midrule


\multirow{3}{*}{\bf 4} & \bf{Ref} & \emph{So \textcolor{red}{from} from that standpoint \textcolor{red}{it's pretty small} it's pretty small}\\

& \bf{Pipe} & \emph{So \hphantom{from} from that standpoint \hphantom{it's pretty small} it's pretty small}\\ 

& {\bf E2E} & \emph{So \hphantom{from} from that standpoint it's pretty small it's pretty small }\\
\midrule

\multirow{3}{*}{\bf 5} & \bf{Ref} & \emph{\textcolor{red}{It's I} I'm sure there's a lot of differences \textcolor{red}{in the way} in the way it's done now and then}\\

& \bf{Pipe} & \emph{\hphantom{It's I} I'm sure there's a lot of differences \hphantom{in the way} in the way it's done now and then }\\ 

& {\bf E2E} & \emph{\hphantom{It's I} I'm sure there's a lot of differences in the way in the way it's done now and then }\\\midrule


\end{tabular}
\end{center}
\caption{Some examples from the SWBD dev set and corresponding transcripts. Ref is the reference transcript which is human-transcribed speech with gold disfluency labels, shown in red. E2E represents the output of the end-to-end Transformer ASR and disfluency removal model. Pipe refers to the output of the pipeline Transformer ASR and ``off-the-shelf'' disfluency detection model.}\label{tab:05} 
\end{table*}

\section{Experiments} \label{sec:ex}
We train our ASR models on two corpora of English conversational telephone speech: \begin{enumerate*}[label=(\roman*)] \item Switchboard-1 Release 2 (SWBD)~\cite{godfrey:93} and \item Fisher Part 1~\cite{cieri:04} and Part 2~\cite{cieri:05}\end{enumerate*}. Switchboard-1 Release 2 is a collection of about 2,400 telephone conversations (260 hours of speech), of which 1,126 conversations were hand-annotated with disfluencies as part of the Penn Treebank Release 3 dataset~\cite{mit:99}, which we refer to as \emph{gold data}. The original release of Switchboard does not contain time-alignment annotations which are required for preparing the ASR training data. Mississippi State University researchers ran a clean-up project on Switchboard-1 Release 2 and produced accurate time alignments\footnote{\url{http://www.openslr.org/5/}} which we use for speech segmentation.

Fisher Part 1 and 2 are a collection of 11,700 telephone conversations (total 2,000 hours of speech), which contain time-aligned transcripts, but no disfluency annotations. To identify the disfluencies in the Fisher data and the portion of the SWBD data with no gold disfluency labels, we use an ``off-the-shelf'' state-of-the-art disfluency detection model\footnote{\url{https://github.com/pariajm/joint-disfluency-detector-and-parser}}~\cite{jam:19}. We call the automatically annotated data \emph{silver data}. The disfluency detection model used to obtain silver data is a BERT-based self-attentive parser that jointly finds a constituency parse tree and detects disfluencies in speech transcripts. Different versions of the parser are available; we use the parser trained on the Penn Treebank Release 3 Switchboard corpus with partial words kept in the data for which they reported an f-score of $94.4$ on the SWBD dev set. We remove all disfluent words (tagged as ``EDITED'' and ``INTJ''), as well as partial words (words tagged ``XX'' and words ending in ``-'') and punctuation from the SWBD and Fisher data. We use the standard data splits for training our models as well as the language models~\cite{char:01}: training data consists of the {\tt sw[23].text} files\footnote{The ``off-the-shelf'' disfluency detection model has been trained on the standard SWBD training split.} and {\tt fe\_03\_$\ast$.txt}, dev data consists of the {\tt sw4[5-9].text} files and test data consists of the {\tt sw4[0-1].text} files.

We consider a pipeline approach as our baseline and apply the ``off-the-shelf'' disfluency detection model to the output of the baseline ASR models. As our evaluation metrics, we report WER, FER and DER for the end-to-end and the pipeline models. Since the goal of an integrated system is to find only the fluent words, we evaluate WER only on fluent words. For calculating FER and DER, we align the output of the integrated models and the output of the pipeline ASR and disfluency detector to the reference transcripts with gold disfluency labels. We report DER results for detecting edited disfluencies, interjections and partial words. In order to have a fair comparison, we report all the results of the paper on the subset of the Switchboard dev and test sets with gold disfluency labels.


\section{Results}
We compare the performance of our integrated ASR and disfluency models (trained on fluent transcripts) to the baseline pipeline models consisting of the ASR models (trained on disfluent transcripts) combined with the ``off-the-shelf'' disfluency detection model. As shown in Table~\ref{tab:02}, the WER of end-to-end models is higher than that of the pipeline models, indicating that word recognition is generally more difficult when the ASR model is trained on disfluent speech and fluent transcripts. 

FER measures the (fluent) word recognition performance of the model while DER reflects how well the model performs in terms of disfluency detection and removal. The baseline ASR models (without disfluency detection) have the lowest error rate on fluent areas (i.e. FER). However, when we apply the ``off-the-shelf'' disfluency detection model on the output of the baseline ASR models, FER significantly increases, indicating that errors made by the ``off-the-self'' disfluency detection model harm the detection of fluent words. The fluent error rate of the end-to-end models is lower than the pipeline models. Comparing the disfluent error rate of the end-to-end and baseline models, we realize that simply training an ASR model on disfluent speech and fluent transcripts significantly decreases the number of disfluencies in the output\footnote{Using the fluent transcripts to train constrains the model not to generate filled pauses such as \emph{uh}, \emph{hm}, \emph{um} and so on.}. However, this is not sufficient for outperforming the baseline pipeline models on detecting and removing disfluencies, indicating that more complex architectures or mechanisms are required for effective end-to-end ASR and disfluency detection. The pipeline models have access to more information (i.e. the annotated disfluencies) than the end-to-end models; however, it is not clear if or how it would improve system performance. Of the three end-to-end models, the Transformer has the best performance on disfluency removal which we speculate is due to the self-attention mechanism which has been previously shown effective in detecting disfluencies in speech transcripts~\citep{tran:19, jam:19, dong:19, wang:19}. We also compare the end-to-end ASR and disfluency removal models with the pipeline ASR and disfluency detection on the Switchboard test set, as demonstrated in Table~\ref{tab:04}.

\renewcommand{\arraystretch}{1.2}
\begin{table}[H]
	\begin{center}
		\begin{tabular}{p{3.36cm}ccc}
			\hline \bf model & \bf WER &  \bf  FER & \bf DER   \\ \hline
        CTC (base)  & 12.4  & 10.2 & 93.5  \\ 
        CTC (pipe)  & 12.4  & 13.5 & 20.2  \\ 
        CTC (E2E)  & 13.6 & 11.1 &  22.6\\ \hline
        Seq2Seq (base) & 8.7 & 7.7 & 95.0 \\
        Seq2Seq (pipe)  & 8.7 & 9.1 & 18.8 \\ 
        Seq2Seq (E2E)  & 10.5 & 8.9 & 21.8 \\ \hline
        Transformer (base)  & 9.5  & 8.5 & 94.6 \\ 
        Transformer (pipe)  & 9.5  & 10.2 & 18.6 \\ 
        Transformer (E2E)  & 11.2  & 9.4 & 20.2  \\ \hline
        Gold Transcripts + DF  & - & 2.2 & 16.8  \\  \hline
		\end{tabular}
	\end{center}
	\caption{Word error rate (WER) with respect to the fluent transcript, fluent error rate (FER) and disfluent error rate (DER) on the \textbf{SWBD dev set}. ``Gold Transcripts + DF'' = the gold transcripts followed by the ``off-the-shelf'' disfluency detector (DF), ``base'' = the baseline ASR (trained on disfluent transcripts), ``pipe'' = the baseline ASR + DF, ``E2E'' = end-to-end ASR and disfluency removal (trained on fluent transcripts).}\label{tab:02} 
\end{table}

\renewcommand{\arraystretch}{1.2}
\begin{table}[H]
	\begin{center}
		\begin{tabular}{p{3.36cm}ccc}
			\hline \bf model & \bf WER &  \bf  FER & \bf DER   \\ \hline
        CTC (base)   & 12.5 & 11.8 & 94.7 \\ 
        CTC (pipe)   & 12.5 & 13.1  & 23.3  \\ 
        CTC (E2E)   & 14.3 & 12.4 & 26.2 \\ \hline
        Seq2Seq (base)  & 11.2 & 10.4 & 95.2 \\
        Seq2Seq (pipe)   & 11.2 & 11.6 & 22.6 \\      
        Seq2Seq (E2E)  & 12.2 & 10.1 & 25.6 \\ \hline

        Transformer (base)  &  11.2 & 10.5  & 95.2 \\ 
        Transformer (pipe)   & 11.2 & 12.1 & 22.2 \\ 
        Transformer (E2E)  & 13.8 & 11.6 & 24.0 \\ \hline
        Gold Transcripts + DF  & - & 2.7 & 17.7   \\  \hline
		\end{tabular}
	\end{center}
	\caption{Word error rate (WER) with respect to the fluent transcripts, fluent error rate (FER) and disfluent error rate (DER) on the \textbf{SWBD test set}. ``Gold Transcripts + DF'' = the gold transcripts followed by the ``off-the-shelf'' disfluency detector (DF), ``base'' = the baseline ASR (trained on disfluent transcripts), ``pipe'' = the baseline ASR + DF, ``E2E'' = end-to-end ASR and disfluency removal (trained on fluent transcripts).}\label{tab:04} 
\end{table}

\begin{table}[H]
	\begin{center}
		\begin{tabular}{lcccc}
			\midrule \bf Model & \bf  Rep. & \bf Cor. & \bf  Res. & \bf All \\ \midrule 
			CTC     & 23.6 & 33.5 & 36.0 & 28.9  \\ 			
			Seq2Seq & 22.5 & 29.5 & 35.1 & 27.1  \\ 
	Transformer     & 22.1 & 25.8 & 35.1 & 25.0   \\
			\midrule
		\end{tabular}
	\end{center}
	\caption{Disfluent error rate (DER) of three end-to-end ASR and disfluency removal models for different types of disfluency on a subset of the SWBD dev set containing $145$ disfluent structures --- including $76$ repetitions (Rep.), $58$ corrections (Cor.) and $11$ restarts (Res.). }\label{tab:03} 
\end{table}

To further investigate the disfluency removal performance of the three end-to-end models, we randomly select 100 sentences from the Switchboard dev set containing disfluencies. We categorize disfluencies into \emph{repetition}, \emph{correction} and \emph{restart} according to the~\citet{shri:94} typology of speech repairs. Repetitions are repairs where the reparandum and repair portions of the disfluency are identical, while corrections are where the reparandum and repairs differ (which are much harder to detect). Restarts are where the speaker abandons a sentence and starts a new one (i.e. the repair is empty). As Table~\ref{tab:03} shows, the end-to-end Transformer model outperforms the other models in detecting all types of disfluency. It particularly has better performance on corrections, which are the more challenging disfluency types in comparison with repetitions.

\subsection{Qualitative Analysis}
We conduct a qualitative analysis on the Switchboard dev set to characterize the disfluencies that the pipeline model cannot detect but the end-to-end model can and vice versa. We provide representative examples in Table~\ref{tab:05}. ASR errors usually lead to disfluency detection errors in the pipeline model (see \#1-3). On the other hand, the end-to-end model sometimes fails at detecting repetitions which are the most common type of disfluency. While the specialized disfluency detector is good at detecting repetitions in speech transcripts, it seems that identifying repetitions in speech signal is non-trivial for the end-to-end model (see \#4 and \#5).

\section{Conclusion}
We showed WER is insufficient for evaluating end-to-end ASR and disfluency removal systems and alternatively introduced two metrics reflecting how well end-to-end systems handle disfluencies. We also showed the disfluency removal performance of end-to-end models is comparable to that of pipeline ASR and specialized high-performance disfluency models. The best end-to-end system uses a Transformer, that's what the best ``off-the-shelf'' disfluency detection system does, too. In the future, we aim to retrain the ``off-the-shelf'' disfluency detector on ASR outputs using cross-validation. It is interesting to investigate how modifying the training loss would affect disfluency detection in end-to-end models. We also intend to augment the end-to-end Transformer model with special mechanisms which have been previously shown effective for disfluency detection in speech transcripts.






\section*{Acknowledgements}
We would like to thank the anonymous reviewers for their insightful comments and suggestions. This research was supported by a Google award through the Natural Language Understanding Focused Program, by a CSIRO's DATA61 Top-up Scholarship, and under the Australian Research Councils Discovery Projects funding scheme (project number DP160102156). 

\bibliography{anthology,emnlp2020}
\bibliographystyle{acl_natbib}

\end{document}